\documentclass[11pt,letterpaper,groupedaddress,superscriptaddress,nofootinbib,floatfix,preprintnumbers,tightenlines]{revtex4}
\usepackage{hyperref,amssymb,amsmath,graphicx,xcolor,cancel}
\usepackage{bm}
\usepackage{placeins}

\def\si{^1 \hskip -0.03in S _0}
\def\siii{^3 \hskip -0.025in S _1}
\def\diii{^3 \hskip -0.025in D _1}

\newcommand{\change}[1]{\textcolor{black}{#1}}

\begin{document}

\begin{figure}[!t]
\vskip -1.1cm
\leftline{
	\includegraphics[width=3.0 cm]{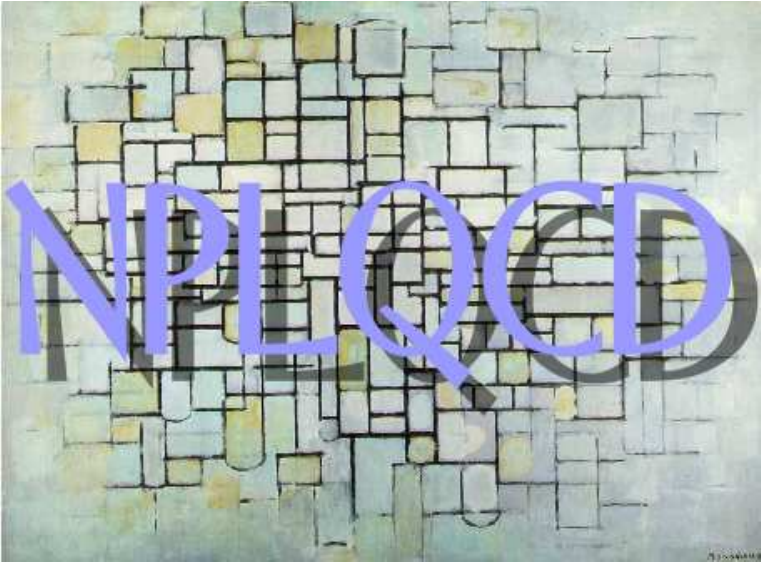}}
\vskip -0.5cm
\end{figure}

\title{Comment on ``Are two nucleons bound in lattice QCD for heavy quark masses? - Sanity check with L\"uscher's finite volume formula -''}

\author{Silas~R.~Beane} 
\affiliation{Department of Physics,
	University of Washington, Box 351560, Seattle, WA 98195, USA}

\author{Emmanuel~Chang}
\noaffiliation

\author{Zohreh Davoudi} \affiliation{
	Center for Theoretical Physics, 
	Massachusetts Institute of Technology, 
	Cambridge, MA 02139, USA}

\author{William Detmold} \affiliation{
	Center for Theoretical Physics, 
	Massachusetts Institute of Technology, 
	Cambridge, MA 02139, USA}

\author{Kostas~Orginos}
\affiliation{Department of Physics, College of William and Mary, Williamsburg,
	VA 23187-8795, USA}
\affiliation{Jefferson Laboratory, 12000 Jefferson Avenue, 
	Newport News, VA 23606, USA}

\author{Assumpta~Parre\~no}
\affiliation{Department of Quantum Physics and Astrophysics and Institute of Cosmos Science ,
	Universitat de Barcelona, Mart\'{\i} Franqu\`es 1, E08028-Spain}

\author{Martin J. Savage}
\affiliation{Institute for Nuclear Theory, University of Washington, Seattle, WA 98195-1550, USA}
\affiliation{Department of Physics,
	University of Washington, Box 351560, Seattle, WA 98195, USA}

\author{Brian~C.~Tiburzi} 
\affiliation{ Department of Physics, The City College of New York, New York, NY 10031, USA }
\affiliation{Graduate School and University Center, The City University of New York, New York, NY 10016, USA }

\author{Phiala E. Shanahan } \affiliation{
	Center for Theoretical Physics, 
	Massachusetts Institute of Technology, 
	Cambridge, MA 02139, USA}

\author{Michael L. Wagman} 
\affiliation{Department of Physics,
	University of Washington, Box 351560, Seattle, WA 98195, USA}
\affiliation{Institute for Nuclear Theory, University of Washington, Seattle, WA 98195-1550, USA}

\author{Frank Winter}
\affiliation{Jefferson Laboratory, 12000 Jefferson Avenue, 
	Newport News, VA 23606, USA}

\collaboration{NPLQCD Collaboration}

\date{\today}

\preprint{INT-PUB-17-016}
\preprint{NT@UW-17-10}
\preprint{MIT-CTP-4909}

\newcommand{\hal}{HAL}

\begin{abstract}
In this comment, we address a number of erroneous discussions and conclusions presented in a recent preprint by the HALQCD collaboration, arXiv:1703.07210~\change{\cite{Iritani:2017rlk}}. 
In particular,
we show that lattice QCD determinations of bound states at  quark masses corresponding to a pion mass of $m_\pi=806$ MeV are robust, and that the extracted phases shifts for these systems pass all of the ``sanity checks'' introduced in arXiv:1703.07210~\change{\cite{Iritani:2017rlk}}.
\end{abstract}

\maketitle


In the last decade, significant progress has been made in 
the study of multi-hadron systems using lattice QCD, with
the first calculations of multi-baryon bound states and their electroweak properties and decays having been performed \cite{Fukugita:1994ve,Beane:2006mx,Ishii:2006ec,Aoki:2008hh,Nemura:2008sp,Yamazaki:2009ua,Aoki:2009ji,Beane:2010hg,Inoue:2010es,Yamazaki:2011nd,Beane:2011iw,Beane:2012vq,Beane:2013br,Inoue:2011ai,Yamazaki:2012hi,HALQCD:2012aa,Beane:2013kca,Beane:2014ora,Beane:2015yha,Detmold:2015daa,Berkowitz:2015eaa,Yamazaki:2015asa,Yamada:2015cra,Chang:2015qxa,Savage:2016kon,Shanahan:2017bgi,Tiburzi:2017iux}. It is imperative that the methods used in these calculations be robust;  investigations such as those of the HALQCD collaboration in Ref.~\cite{Iritani:2017rlk} are vital provided they are carried out correctly. However, as we show in detail, many of the conclusions reached in Ref.~\cite{Iritani:2017rlk} (henceforth referred to as \hal), that cast doubt on the validity of multi-baryon calculations, are incorrect. Since we have recently refined one of the analyses that is criticized in \hal{}, we focus our attention on the conclusions drawn regarding this case in particular, see Ref.~\cite{Wagman:2017tmp}. 

The central point addressed by \hal{} is whether there exist bound states in the $\si$ and $\siii$ two-nucleon channels at heavy quark masses. Three independent groups have analysed lattice QCD calculations at quark masses corresponding to a heavy pion mass of $\sim800$ MeV (one set of calculations used quenched QCD) and found that there are bound states in these channels. Each of these groups has concluded this by extracting energies from two-point correlation functions (with the quantum numbers of interest) at two or more lattice volumes and demonstrating, through extrapolations based on the finite-volume formalism of L\"uscher \cite{Luscher:1986pf,Luscher:1990ux}, that these energies correspond to an infinite-volume state that is below the two-particle threshold and is hence a bound state. Each group has used different technical approaches, and all are in reasonable agreement given the uncertainties that are reported. The HALQCD collaboration has also investigated these two-particle channels using a method (also based on the work of L\"uscher \cite{Luscher:1986pf,Luscher:1990ux}) that involves constructing Bethe-Salpeter wavefunctions, but do not find evidence for bound states in these channels \cite{Ishii:2006ec,Aoki:2008hh,Aoki:2009ji,HALQCD:2012aa,Inoue:2011ai}.\footnote{In the $\Lambda\Lambda$ channel, the HALQCD approach does indicate a bound state, but the binding energy is found to be significantly different from that determined by extrapolating finite-volume energy levels \cite{Beane:2012vq}.} We note, however, that the HALQCD method introduces unquantified systematic effects as discussed in, e.g., Refs.~\cite{Detmold:2007wk,Beane:2010em,Detmold:2015jda} and the nuclear physics overview talks in recent proceedings of the International Symposium on Lattice Field Theory \cite{Walker-Loud:2014iea,Yamazaki:2015nka,Savage:2016egr}). Here, we focus our criticisms of \hal{} on several specific points.

\subsubsection{Misinterpretation of energies and source independence} 
\begin{figure}[!t]
	\includegraphics[width=0.95\columnwidth]{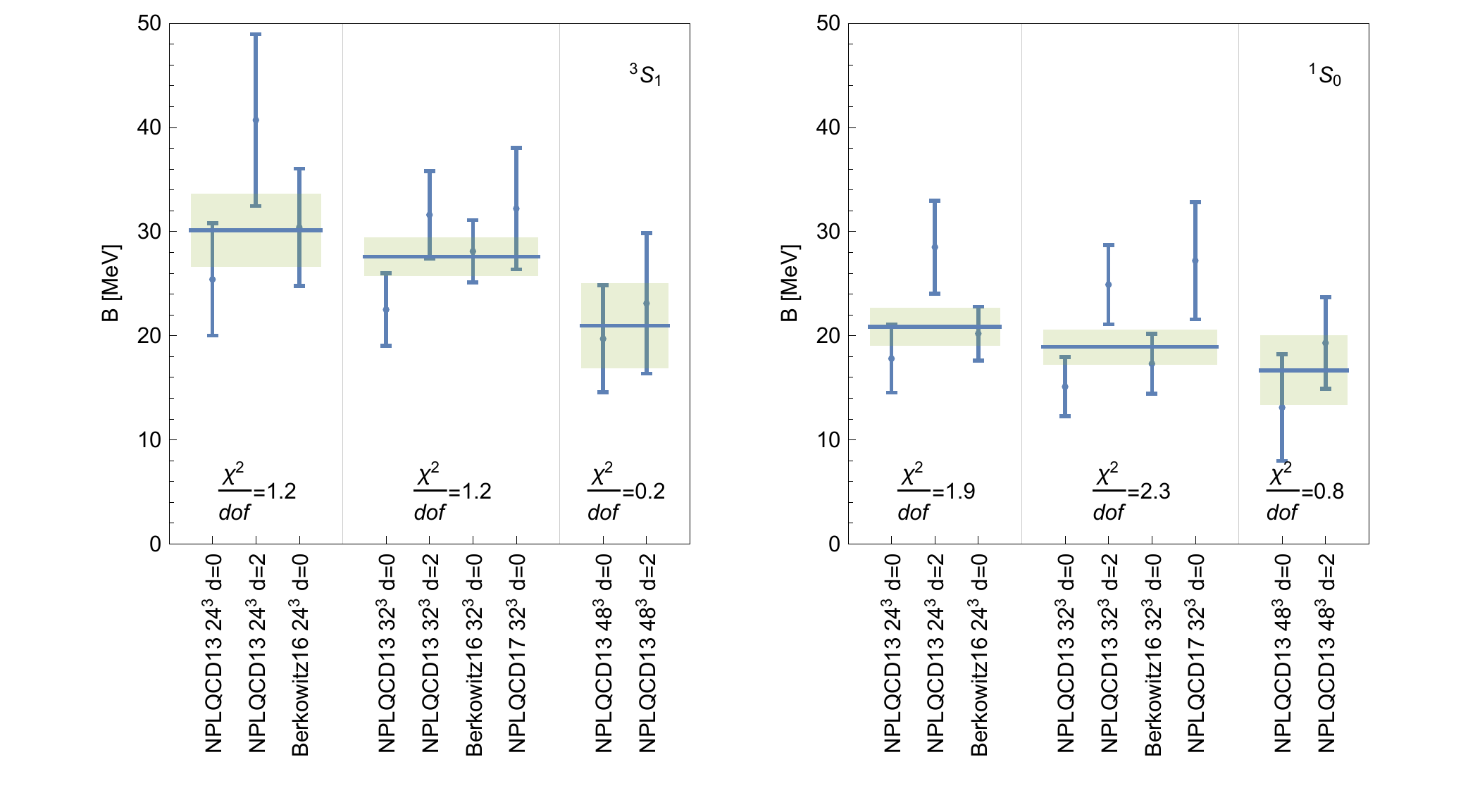}
	\caption[.]{Binding energies of the $\siii$ and $\si$ ground states at $m_\pi= 806$ MeV found in the literature: NPLQCD13 \cite{Beane:2012vq}, Berkowitz16 \cite{Berkowitz:2015eaa}, and NPLQCD17 \cite{Savage:2016kon,Shanahan:2017bgi,Tiburzi:2017iux} ($d=0$ and $d=2$ refer to the magnitude of the centre-of-mass momentum used in the calculations in units of $2\pi/L$). The three regions in each panel correspond to three different volumes: $L=24$, 32, and 48 from left to right. Uncertainties listed in the original references are combined in quadrature. The horizontal lines and shaded bands represent the central value and one standard deviation bands from uncorrelated fits, respectively.}
	\label{fig:binding}
\end{figure}

Figure 2 of \hal{} contains a compilation of results for the ground states of the $\si$ and $\siii$ two-nucleon channels. Unfortunately the figure includes a second state 
from Ref.~\cite{Berkowitz:2015eaa} that the authors of Ref.~\cite{Berkowitz:2015eaa} explicitly indicate is not the ground state, and reporting it as such is a significant error on which many of the invalid arguments of HAL are based.\footnote{Whether the quoted value for the second energy in Ref.~\cite{Berkowitz:2015eaa} is a true estimate of an excited-state energy is a question for future discussion. However for the ground states, all results unambiguously agree.} There is a small scatter in the remaining results that is due to statistical fluctuations, discretisation artifacts and exponentially-small residual finite-volume effects, but, taken as a whole, there is no inconsistency in these results. In addition, a further recent study of axial-current matrix elements using a different set of interpolators \cite{Savage:2016kon,Shanahan:2017bgi,Tiburzi:2017iux} (denoted in Fig.~\ref{fig:binding} by NPLQCD17) also finds a consistent negatively-shifted energy on the $32^3\times48$ ensemble used in this comparison. 
Figure 2 in \hal{} also fails to include the energies extracted in Ref.~\cite{Beane:2012vq} on the largest volume, which dominate the extraction of the binding energy. Without the results from this large volume, the confidence in the binding energy in Ref.~\cite{Beane:2012vq} would be significantly diminished.  It is therefore vital that this information be included in any discussion of these results. Figure \ref{fig:binding} below shows a (corrected) summary of the energy levels extracted for the ground states of the $\si$ and $\siii$ two-nucleon systems in different volumes that are published in the literature 
 at this particular quark mass.  No significant interpolator dependence is observed, as is indicated by simple fits to the reported results for each volume,  with all these fits having acceptable values of $\chi^2$ per degree of 
freedom.
Figure 13 of \hal{} is also erroneously described as indicating that scattering state results are not source independent. The results show three energy levels where different interpolating operators are consistent within one standard deviation, and one energy level that differs at two standard deviations. This indicates broad agreement within the reported uncertainties and, contrary to statements in HAL, does not provide a sound statistical basis for a claim of inconsistency. 

In summary, comparison of results from the different interpolators in Refs. \cite{Beane:2012vq,Beane:2013br,Berkowitz:2015eaa,Tiburzi:2017iux} shows that both bound and scattering-state energy levels are source-independent within reported uncertainties. This is contrary to the claims in \hal{}.

\subsubsection{Volume scaling of energies}
\begin{figure}[!t]
	\includegraphics[width=0.8\columnwidth]{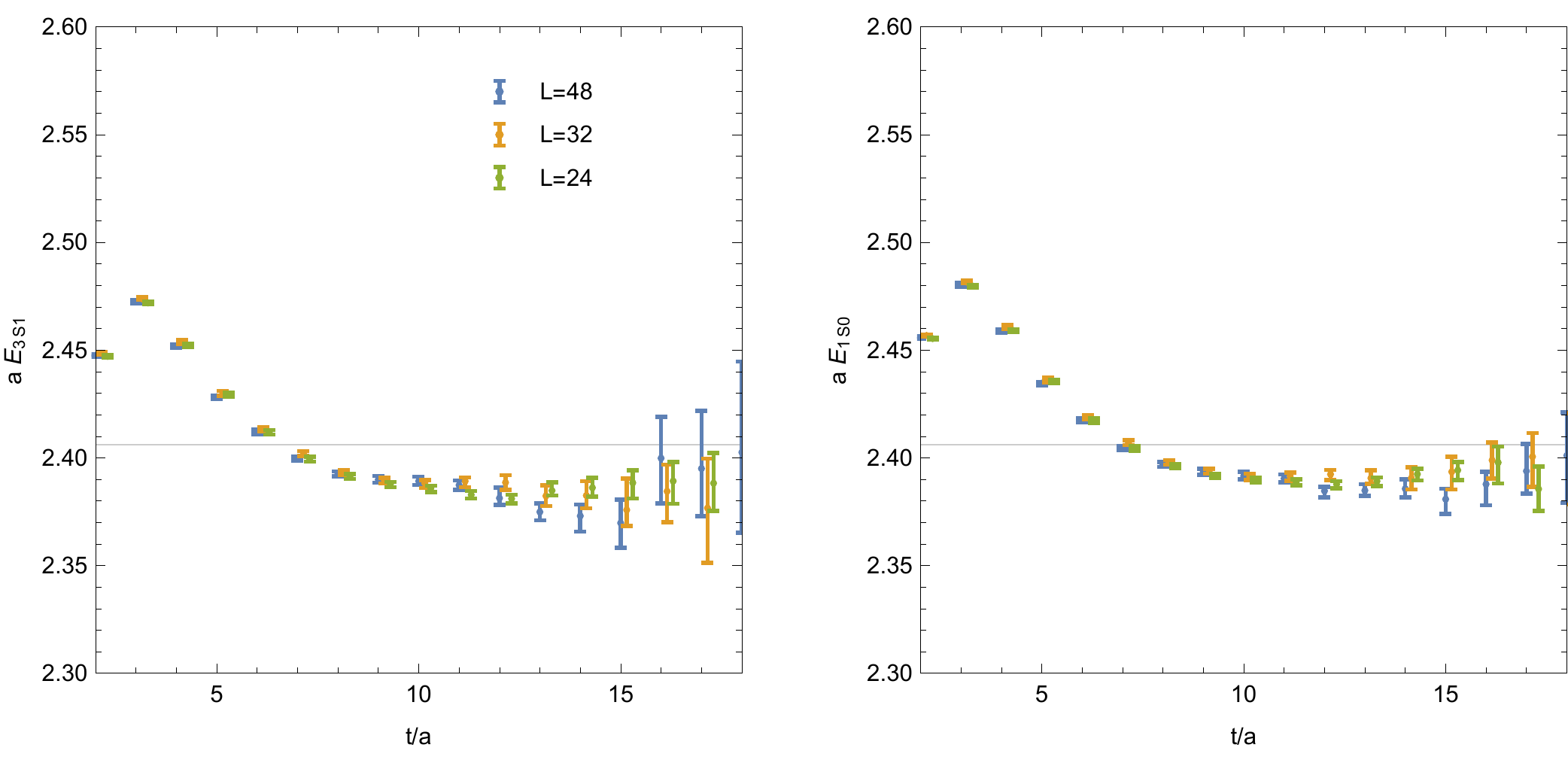}
	\caption[.]{The effective mass plots associated with the $d=0$ smeared-point correlators in the $L=24$, 32, and 48 ensembles of Ref.~\cite{Beane:2012vq,Beane:2013br}. The left(right) panel shows the $\siii$($\si$) channel. Quantities are expressed in lattice units. The horizontal grey line marks the infinite-volume energy of two non-interacting nucleons.}
	\label{fig:eff}
\end{figure}

The authors of HAL claim that the single-exponential behaviour found in our work, Refs.~\cite{Beane:2012vq,Beane:2013br}, and in that of Ref.~\cite{Berkowitz:2015eaa}, is a ``mirage'' arising from the cancellation of two or more scattering eigenstates\footnote{The scattering states are loosely used here to denote states in a finite volume that correspond to the continuum states of infinite volume.} contributing to the correlation functions with opposite signs (see Ref.~\cite{Iritani:2016jie} for elaborations on possible ``mirage'' plateaus). This interpretation of the negatively-shifted states in these works is exceedingly unlikely, however, as such cancellation would need to occur in an almost identical way for multiple different volumes. For each of the different analyses of the 806 MeV ensembles in Fig.~\ref{fig:binding} (NPLQCD2013 $d=0$, NPLQCD2013 $d=2$ and Berkowitz2016 $d=0$), identical sources and sinks were used in each of three volumes (two volumes in the case of Berkowitz2016). Scattering-state eigenenergies  necessarily change significantly with volume, having power-law dependence as dictated by the L\"uscher quantisation condition. While it is possible that, in a given volume, a correlator for a particular source-sink interpolator combination could exhibit a cancellation between contributions of two scattering states that produces an energy level below threshold, it is very unlikely that the cancellation would persist in different volumes as the scattering-state eigenenergies change significantly with volume. As shown in Fig.~\ref{fig:eff}, for example, the volume-independent interpolators used in Ref.~\cite{Beane:2012vq,Beane:2013br} produce energy levels in the three different volumes that are statistically indistinguishable, and even the approach to single-exponential behaviour does not depend on volume. The figure shows the effective masses of the smeared-point correlation functions, but the same features are seen in all other source-sink interpolator combinations that are studied. This rules out the possibility that the negatively-shifted signals are caused by cancellations between scattering states. The largest volume used in our works \cite{Beane:2012vq,Beane:2013br} makes this an extremely robust statement as the spatial volumes from which we draw these conclusions vary by a factor of eight.

\subsubsection{Consistency of Effective Range Expansion (``\hal{}  Sanity Check (i)'')} 
\begin{figure}[!t]
\includegraphics[width=0.8\columnwidth]{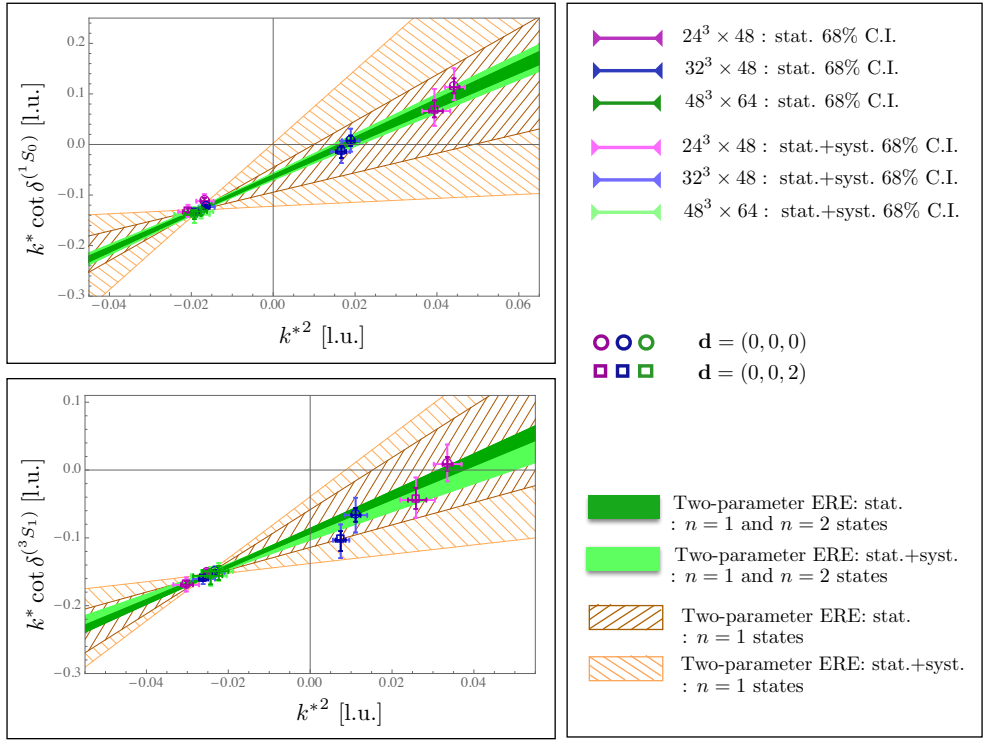}
\caption[.]{$k^*\cot \delta$ vs. the square of the centre-of-mass momentum of two baryons, ${k^*}^2$, along with the bands representing fits to two-parameter EREs obtained from i) only the ground states ($n=1$) and ii) from both the ground states ($n=1$) and the first excited states ($n=2$). The plots show the consistency of the ERE between negative and positive ${k^*}^2$ regions in both the $\si$ and $\siii$ channels. These results are from our recent re-analysis of these ensembles~\cite{Wagman:2017tmp}, and are consistent with the initial analysis~\cite{Beane:2012vq,Beane:2013br}, with the mean values in agreement within one standard deviation as defined by the combined (statistical and systematic) uncertainties of each result. Quantities are expressed in lattice units (l.u.).}
\label{fig:ERE-n1-n2}
\end{figure}

If the effective range expansion (ERE) is a valid parametrization of the scattering amplitude at low energies, the analyticity of the amplitude as a function of the centre-of-mass energy implies that the  ERE obtained from states with positively-shifted energies (${k^*}^2>0$, where $k^*$ is the centre-of-mass interaction momentum) must be consistent with that obtained from states with negatively-shifted energies (${k^*}^2<0$). Although \hal{} finds that the NPLQCD results pass this test, we  demonstrate how robust the results in Refs.~\cite{Beane:2012vq,Beane:2013br} are in this regard through the plots presented in Fig. \ref{fig:ERE-n1-n2}. This figure shows fits to the ERE using both ground states ($n=1$) and first excited states ($n=2$) (color-shaded bands). These are overlaid on ERE fits using only the ground states (hashed bands). The two sets of bands are fully consistent with each other, proving that this check is unambiguously passed. The same feature is seen for three-parameter ERE fits, with significantly larger uncertainty bands (see also Ref.~\cite{Wagman:2017tmp}).\footnote{Our analysis of two-nucleon correlation functions generated from these ensembles of gauge-field configurations has been recently refined in a comprehensive re-analysis~\cite{Wagman:2017tmp},  including results at additional kinematic points. This new analysis has been used in obtaining the results shown in Figs.~\ref{fig:ERE-n1-n2} and \ref{fig:ERE-tangent}.  All of the energies extracted from the three lattice volumes, and the binding energies and ERE parameters subsequently obtained, are in agreement with our previous results; i.e., the differences in the mean values of the results from the previous and the new analyses are within one standard deviation as defined by the (statistical and systematic) uncertainties of the results combined in quadrature~\cite{Beane:2012vq,Beane:2013br}.}  
The difference in the size of uncertainties in the phase shift between the fits with and without the $n=2$ data shows that conclusions about the behaviour and/or validity of the ERE for datasets only near the bound-state pole are likely subject to significant uncertainties. We note that scattering parameters extracted in the region near ${k^*}^2=0$ from a linear ERE will in general differ from those determined in the vicinity of a bound-state pole due to higher order terms in the ERE. Indeed, it is known that in nature, the ERE of the $\siii$ phase shift around ${k^*}^2=0$ and around the deuteron pole are different (albeit slightly) \cite{deSwart:1995ui}.
%

\subsubsection{Residue of the S-matrix at the bound-state pole (``\hal{} Sanity check (iii)'')} 
\begin{figure}[h!]
\includegraphics[width=0.8\columnwidth]{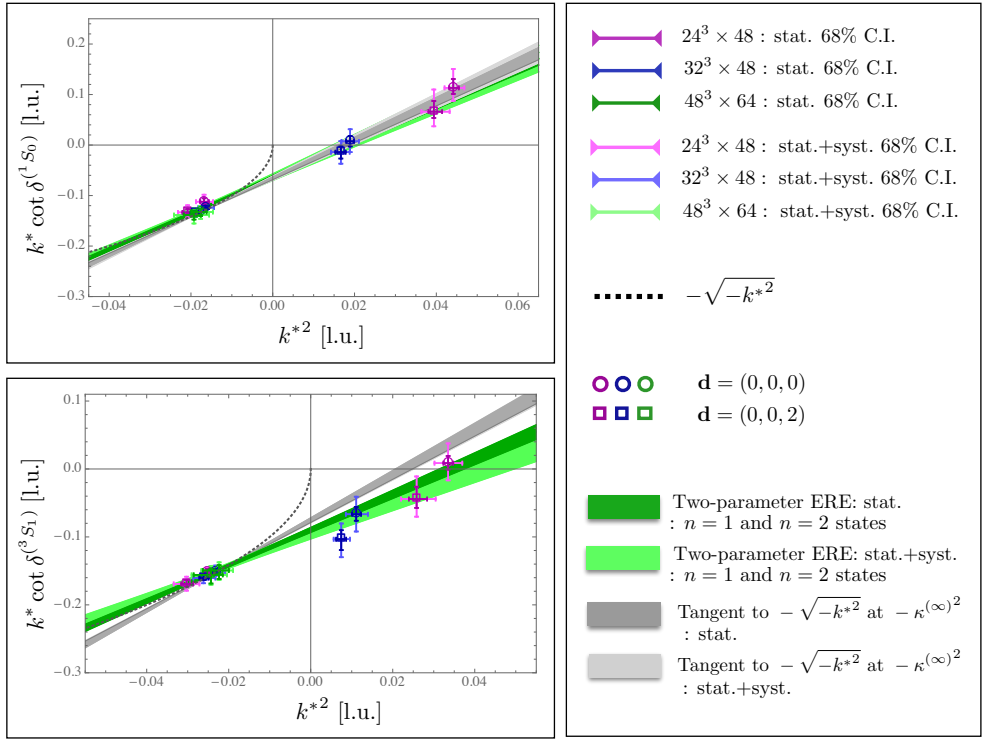}
\caption[.]{
The two-parameter ERE is compared with the tangents to the $-\sqrt{-{k^*}^2}$ curve at values of ${k^*}^2=-{\kappa^{(\infty)}}^2$. 
The plots show that all the identified energy  eigenstates in this work are consistent with the criterion in Eq. (\ref{eq:slope}) within uncertainties. 
These results are from our recent re-analysis of these ensembles~\cite{Wagman:2017tmp}, and are consistent with the initial analysis~\cite{Beane:2012vq,Beane:2013br}, 
with the mean values in agreement within one standard deviation as defined by the combined (statistical and systematic) uncertainties of each result. 
Quantities are expressed in lattice units (l.u.).
\change{
As the two shaded regions in the $\si$ channel are somewhat similar, one is partially obscured by the other in the upper panel.
}
}
\label{fig:ERE-tangent}
\end{figure}

\begin{figure}[h!]
\includegraphics[width=0.8\columnwidth]{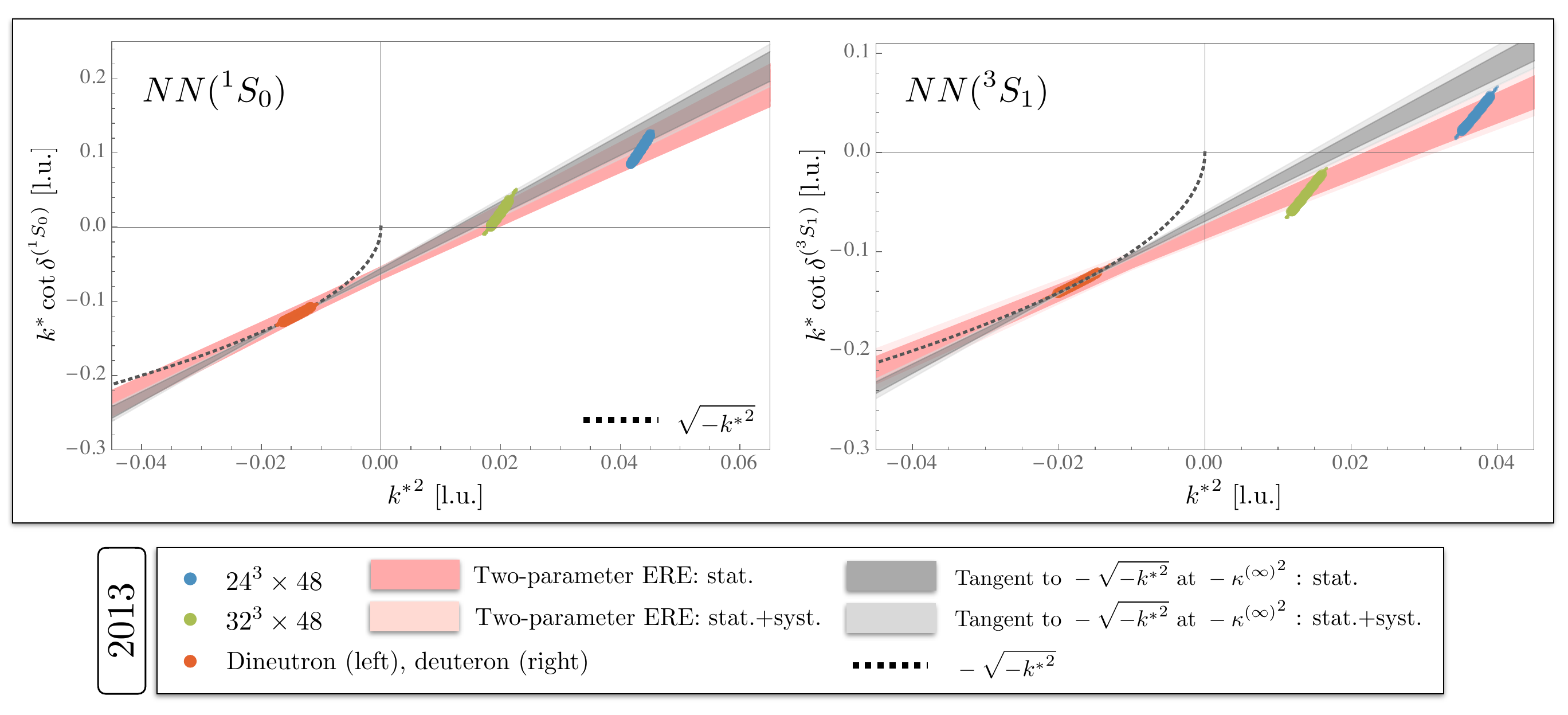}
\caption[.]{
\change{
\change{
Two-parameter ERE fits determined from the 2013 analysis of the 2013 correlation functions~\cite{Beane:2012vq,Beane:2013br} (salmon shaded regions)
are compared with the tangents to the $-\sqrt{-{k^*}^2}$ curve at values of ${k^*}^2=-{\kappa^{(\infty)}}^2$ (grey shaded regions). 
As the two shaded regions in the $\si$ channel are somewhat similar, one is partially obscured by the other in the left panel.
The dashed-black line corresponds to $-\sqrt{-{k^*}^2}$.
Quantities are expressed in lattice units (l.u.).
}
}
}
\label{fig:tangent13}
\end{figure}

The sign of the residue of the S-matrix at the bound-state pole is fixed. This requirement leads to the following condition on  $k^*\cot \delta$ :
\begin{eqnarray}
	\left. \frac{d}{d{k^*}^2}(k^*\cot \delta+\sqrt{-{k^*}^2}) \right |_{{k^*}^2=-{\kappa^{(\infty)}}^2} < 0,
	\label{eq:slope}
\end{eqnarray}
where $\kappa^{(\infty)}$ is the infinite-volume binding momentum. 
As is seen from Fig.~\ref{fig:ERE-tangent}, which displays the results of 
\change{
the 2017 refined analysis~\cite{Wagman:2017tmp} of the correlation functions analyzed in
Refs.~\cite{Beane:2012vq,Beane:2013br}, 
the slope of the two-parameter ERE fit to the $k^*\cot \delta$ function (colored regions) is
less than the slope of $-\sqrt{-{k^*}^2}$ (grey regions) 
at the corresponding bound-state pole in both channels.
In the $\si$ channel the difference is at the $1\sigma$ level, while the difference is more than $3\sigma$ in the coupled $\siii$-$\diii$ channels.
} 
The uncertainty in the tangent line to the $-\sqrt{-{k^*}^2}$ function at ${k^*}^2=-{\kappa^{(\infty)}}^2$ arises from the uncertainty in 
the values of $\kappa^{(\infty)}$ (see also Ref.~\cite{Wagman:2017tmp}). A similar conclusion can be drawn from three-parameter ERE fits. 
\change{
For the sake of clarity, the two-parameter ERE fits to the results of only the 2013 analysis  
of the same correlation functions  are shown in Fig.~\ref{fig:tangent13},
and are seen to be  consistent with the criterion in Eq. (\ref{eq:slope}) as well.}
\begin{figure}[h!]
\includegraphics[width=0.8\columnwidth]{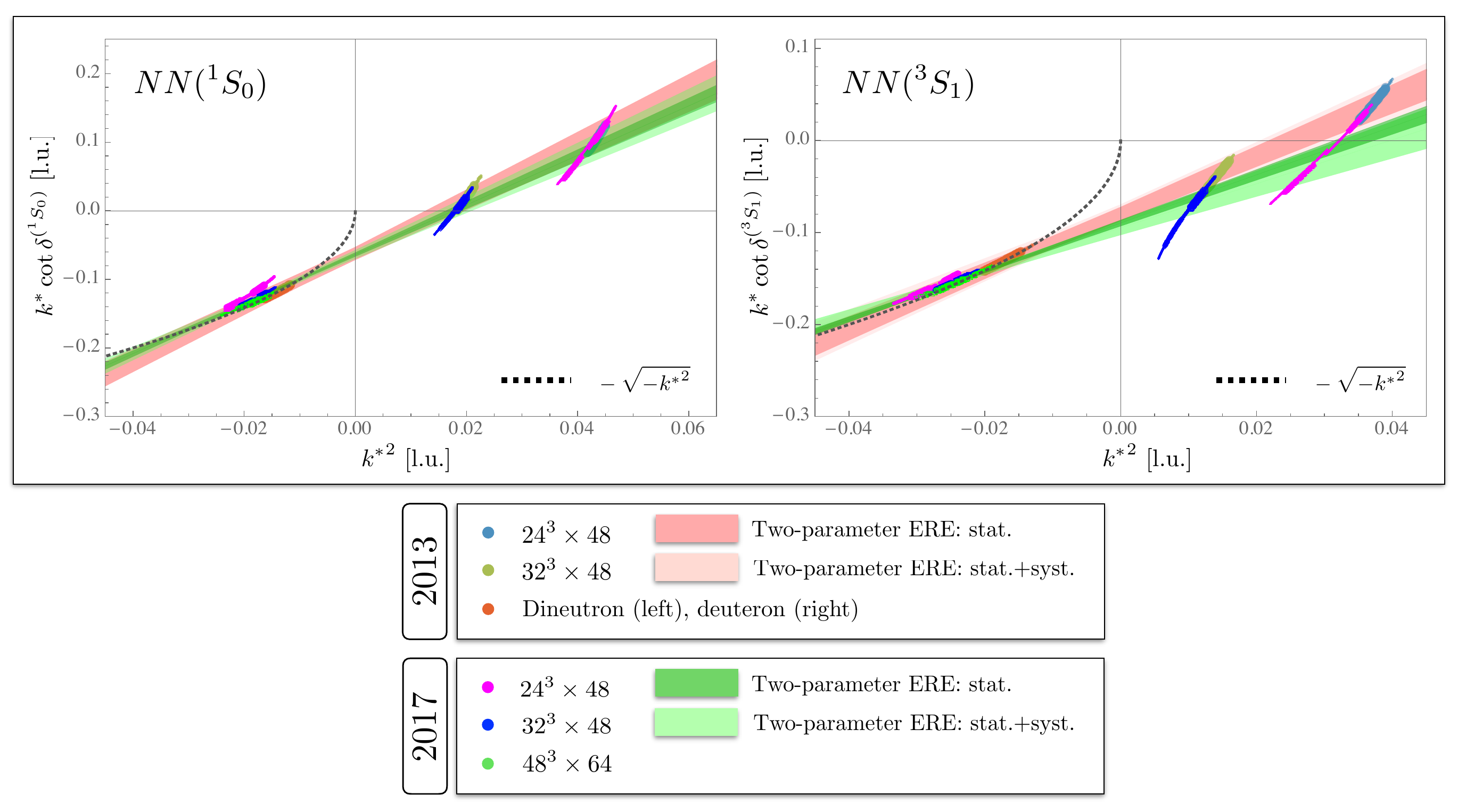}
\caption[.]{
\change{
A comparison between the two-parameter ERE fits to the 2013~\cite{Beane:2012vq,Beane:2013br} 
(salmon shaded regions)
and 2017~\cite{Wagman:2017tmp} 
analyses of the 2013 correlation functions~\cite{Beane:2012vq,Beane:2013br} (green shaded regions).
The dashed-black line corresponds to $-\sqrt{-{k^*}^2}$.
As the two shaded regions in the $\si$ channel are somewhat similar, one is partially obscured by the other in the left panel.
Quantities are expressed in lattice units (l.u.).
}
}
\label{fig:tangent1317}
\end{figure}
\change{
For comparison, 
in Fig.~\ref{fig:tangent1317} we show the results of two-parameter ERE fits obtained in the 2013 analysis~\cite{Beane:2012vq,Beane:2013br}
and in the  2017 refined analysis of the same correlation functions~\cite{Wagman:2017tmp}.
Both analyses of these channels yield results that are consistent with each other and
with the criterion in Eq. (\ref{eq:slope}) within the uncertainties of the calculations, thus passing check (iii). 
}

\subsubsection*{Discussion}
Given the discussion above, 
the NPLQCD results presented in the ``NPL2013'' row of Table IV of the published version of 
\hal{}~\cite{Iritani:2017rlk}, reproduced below,  
\begin{table}[!h]
	\begin{ruledtabular}
		\begin{tabular}{|c||c|c|c|c|c|c|c|c|}
			& \multicolumn{4}{c|}{$NN(^1S_0)$} & \multicolumn{4}{c|}{$NN(^3S_1)$} \\ 
			\hline
			Data & Source & \multicolumn{3}{c|}{Sanity check} & Source & \multicolumn{3}{c|}{Sanity check} \\
			& independence & (i) & (ii) & (iii) & independence  & (i)  & (ii) & (iii) \\
			\hline\hline
			NPL2013 [28,29]   & No       &        *         &      *      &  No    &   No         &    *     &      *        &  ?         \\
		\end{tabular}
	\end{ruledtabular}\\

\begin{flushleft}
	\noindent should be replaced by 
	\change{
	(with reference numbers changed to relate to the present bibliography) 
	}
\end{flushleft}

	\begin{ruledtabular}
		\begin{tabular}{|c||c|c|c|c|c|c|c|c|}
			& \multicolumn{4}{c|}{$NN(^1S_0)$} & \multicolumn{4}{c|}{$NN(^3S_1)$} \\ 
			\hline
			Data & Source & \multicolumn{3}{c|}{Sanity check} & Source & \multicolumn{3}{c|}{Sanity check} \\
			& independence & (i) & (ii) & (iii) & independence  & (i)  & (ii) & (iii) \\
			\hline\hline
			\change{ NPL2013~\cite{Beane:2012vq, Beane:2013br}   }& Yes           &      Passed        &     Passed     & Passed    &   Yes        &   Passed     &      Passed       &  Passed         \\
			\change{NPL2017~\cite{Wagman:2017tmp}  of NPL2013 }& Yes            &       Passed       &    Passed      & Passed    &   Yes        &   Passed     &      Passed       &  Passed         \\
		\end{tabular}
	\end{ruledtabular}
\end{table}\\
\FloatBarrier
\noindent where we have taken the liberty of changing the notation (in their published version) 
used to indicate passing a ``sanity check'' in \hal{} from a `` * '' entry to ``Passed''. 
We are currently
revisiting the other NPLQCD analyses discussed in \hal{}. Ref.~\cite{Yamazaki:2017euu} refutes the \hal{} criticisms of 
source-dependence leveled at the works of the PACS-CS collaboration \cite{Yamazaki:2009ua}. Ref.~\cite{Savage:2016egr} provides a summary of the evidence for the validity of ground-state identifications in two-nucleon systems. With the robust conclusion of the existence of bound states reached by independent groups, and argued in this Comment, the systematic uncertainties of the potential method used by the HALQCD collaboration requires further investigation to better understand the origin of its failure to identify two-nucleon bound states.

\begin{acknowledgments}
	SRB was partially supported by NSF continuing grant number PHY1206498 and by the U.S. Department of Energy through grant
number DE-SC001347.  
EC was supported in part by the USQCD SciDAC project, the U.S. Department of Energy through 
grant number DE-SC00-10337,  and by U.S. Department of Energy grant number DE-FG02-00ER41132.
ZD, WD and PES were partly supported by  U.S. Department of Energy Early Career Research Award DE-SC0010495 and grant number DE-SC0011090.
KO was partially supported by the U.S. Department of Energy through grant
number DE- FG02-04ER41302 and through contract number DE-AC05-06OR23177
under which JSA operates the Thomas Jefferson National Accelerator Facility.  
A.P. is partially supported by the Spanish Ministerio de Economia y Competitividad (MINECO) under the project MDM-2014-0369 of ICCUB (Unidad de Excelencia ’María de Maeztu’), and, with additional European FEDER funds, under the contract FIS2014-54762-P, by the Generalitat de Catalunya contract 2014SGR-401, and by the Spanish Excellence Network 
on Hadronic Physics FIS2014-57026-REDT. 
MJS was supported  by DOE grant number~DE-FG02-00ER41132, and  in part by the USQCD SciDAC project, 
the U.S. Department of Energy through grant number DE-SC00-10337.	
BCT was supported in part by the U.S. National Science Foundation, under grant
number PHY15-15738. 
MLW was supported  in part by DOE grant number~DE-FG02-00ER41132.
FW was partially supported through the USQCD Scientific Discovery through Advanced Computing (SciDAC) project 
funded by U.S. Department of Energy, Office of Science, Offices of Advanced Scientific Computing Research, 
Nuclear Physics and High Energy Physics and by the U.S. Department of Energy, Office of Science, Office of Nuclear Physics under contract DE-AC05-06OR23177.
\end{acknowledgments}
%

\bibliography{bibi.bib}
\end{document}